\documentstyle[aps]{revtex}

\begin{document}

\author{M.A. Khanbekyan}

\title{Generation of coherent superposition of Fock states in a
  cavity. Entanglement of atomic coherent states.}

\address{Institute for Physical Research , Armenian National Academy of
  Sciences,\\
378410, Ashtarak-2, Armenia.}
\maketitle

\begin{abstract} 
A scheme for preparation of coherent superposition of Fock states of
electromagnetic field is constructed. The superposition state is created
inside the cavity via a strong interaction of a four-level atom with quantum
field of the cavity and classic laser fields. We demonstrate the possibility
to create desired arbitrary superposition of the cavity Fock states just by
changing the relative delay of the fields. Then, as another application of
our model, we study a means of creating entanglement of neutral four-level
atoms using different sequences of interactions with the cavity and laser
fields.
\end{abstract}

\section{ INTRODUCTION}
The past few years have seen an upsurge of interest in the creation of
non-classical states of electromagnetic field, such as squeezed states, Fock
states \cite{Knight}. These states have no classical analogies and exhibit
particular statistical properties. Relevant examples are superpositions of
Fock states, which can have applications in quantum computing \cite{Cirac}.

Of particular interest in the generation of quantum states of light fields is
that associated with cavity QED, in which beams of atoms strongly interact
with a single cavity electromagnetic mode. It is assumed that atom-cavity
coupling strengths much larger than dissipative rates.

Yet another class of theory and experiments relevant to quantum communication
and computation procedures is generation of entangled states
\cite{Ekert}. Entangled states are interesting because they exhibit
correlations that have no classical analog.

A variety of proposals have been put forward for the generation of Fock
states in a cavity. The most simple system is a single two-level atom coupled
with cavity-mode field. Theoretical studies indicate the possibility to
create and arbitrary prescribed superposition of Fock states in a cavity via
two-level atom interacting with quantum and classical fields \cite{Law}. This
model requires precise control of time sequences of amplitudes and phases of
atom-cavity and atom-laser coupling fields. In a similar vein, Domokos et
al. \cite{Domokos} demonstrated an experimental scheme for the generation of
Fock states in a cavity. Initially excited Rydberg two-level atoms pass
through and interact with two quantized field modes of the microwave
cavity. Specifically, the first mode is a photon reservoir prepared in a
coherent state and the next one stores the Fock states.

Finally, a Fock states preparation scheme was suggested by Parkins et al.
\cite{Parkins}, in which a beam of $\Lambda$-system atoms interact with a
single quantized field mode of cavity and then with classical laser
field. The initial eigenstate of the atom-cavity system, which corresponds to
the atom being in first ground state, adiabatically evolves into a final
eigenstate, which corresponds to the atom in the ground state and the cavity
in a Fock state. So, if initially there is no photon in the cavity, it is 
created one-photon Fock state. Because of the system evolution corresponds to
the dark state, which has no contribution from excited atomic state,
spontaneous emission does not figure in the dynamics of the system.    

A scheme for preparation of pairs of massive atoms in an entangled
state was predicted by Cirac and Zoller \cite{CZ} and then observed by Hagley
et al. \cite{Hagley}.  The states were produced by interaction with a high-Q
cavity. The technique of Gerry and Grobe \cite{GG} for generation of an
entangled state assumes a large number of two-level atoms confined in a cavity.
 
In the present paper we describe in detail a method for generation of
coherent superposition of Fock states in a cavity and demonstrate its
feasibility by analytic and numerical calculations. Our scheme requires
the passage of a four-level atom through overlapping two classical laser
fields and quantum cavity-mode field. The first laser field of frequency
$\omega _{1}$ couples the atomic state $\left| \psi_1 \right\rangle $ with the
excited state $\left| \psi_2\right\rangle $. The second laser field of frequency
$\omega _{2}$ couples the state $\left| \psi_3\right\rangle $ with the excited state
$\left| \psi_2\right\rangle$. The quantum cavity-mode field couples the states
$\left| \psi_4\right\rangle $ and $\left| \psi_2\right\rangle $. Figure~\ref{fourl}
illustrates the connections. Unanyan et al. \cite{Unanyan} have shown that
using appropriate delay and sequence of interaction pulses one can create
robust coherent superposition of atomic states. Our scheme is based on those
results. We show that, if one of the pulses is the quantum field of the
cavity, a coherent superposition of quantum states is generated in the
cavity. In fact, our scheme requires the following sequence of interactions
with the atom (see Fig.~\ref{pulses}): the cavity-mode field is turned on
before the first laser field arrives and is turned off after $V_1$
ceases. Further, the second laser fields $V_2$ is turned on right after laser
pulse and turned off the last. Of most interest for us is the case when the
cavity quantum field is initially in the vacuum state
$\left|n=0\right\rangle$ and the passage of the atom through the interaction
region yields a superposition of zero-photon $\left|n=0\right\rangle$ and
one-photon $\left|n=1\right\rangle$ states. By changing the relative delay of
the pulses, one obtains arbitrary superposition of cavity quantum field
states.

Further, we consider two four-level atoms, which are initially in the states
$\left| \psi_1 \right\rangle $ and  $\left| \psi_4 \right\rangle $
correspondingly. At early times the first atom interacts with the sequence of
cavity field and equal laser fields ($V_1 (t) = V_2 (t)$). At late
times, the second atom passes through the reversed configuration of pulses,
i.e. equal laser pulses and then cavity field (see Fig.~\ref{entang}). We
will show that these sequences of interactions produce an entanglement of two atoms.

The outline of the rest of the paper is the following: In Sec. II we describe
the resonant interaction of a four-level atom with two laser fields and
quantized cavity field mode. Sec. III is denoted to demonstration of the method
in the case of delayed laser pulses. In Sec. IV we present a scheme for
preparation of entanglement of two atoms. We summarize main results of the
paper in Sec. V.
                        
\section{ ADIABATIC INTERACTION OF FOUR-LEVEL ATOM WITH TWO LASER FIELDS AND
  QUANTIZED CAVITY FIELD MODE} 
We consider a single four-level atom with three ground states $\left|
  1\right\rangle $, $\left|3\right\rangle $, $\left|4\right\rangle $ coupled
to an excited state $\left|2\right\rangle $ via, respectively, two classical
laser fields $V_{1,2}$ of frequency $\omega _{1,2}$ and a cavity-mode field
of frequency $\omega $ (see Fig.~\ref{config}). The dynamics of the system is
described by the Schr\"{o}dinger equation 
     
\begin{equation}
i\hbar \frac{d }{d t}\left| \Psi (t) \right\rangle =H (t)\left| \Psi (t) \right\rangle ,  
\label{Schrodinger}
\end{equation}
with the Hamiltonian
\begin{eqnarray}
H(t) &=&H_{at}+\hbar \omega c^{+}c +\big( V_1(t)\left|1\right\rangle
\left\langle 2\right| + H.c.\big)+\big(
V_2(t)\left|3\right\rangle\left\langle 2\right| + H.c.\big) +
\hbar\big(c^{+}\beta (t) \left|4\right\rangle\left\langle 2\right| +
H.c.\big),
\label{Hamiltonian}
\end{eqnarray}
where $c$ is the annihilation operator for the cavity-mode, $\beta (t)$ is
the atom-cavity-mode coupling strength and $V_{1,2}(t)=W_{1,2}e^{i\omega
  _{1,2}t}+W_{1,2}^{*}e^{-i\omega_{1,2}t}$. The time dependence of $W_{1,2}$
and $\beta (t)$ is provided by the motion of the atom through the interaction
region.

We express the solution of Schr\"{o}dinger equation (\ref{Schrodinger}) as a
superposition of atomic basis states:
\begin{equation}
\left| \Psi (t) \right\rangle =\sum_{i} a_{i} (t)\left|
  \psi_{i}\right\rangle,   
\label{basis}
\end{equation}
here $\left| \psi_{i}\right\rangle$ are the eigenvectors of the free atomic
Hamiltonian
\begin{equation}
H_{at}\left| \psi_{i}\right\rangle=\varepsilon _{i}\left|
  \psi_{i}\right\rangle, \quad (i=1,2,3,4).  
\label{basis1}
\end{equation} 
Further, we transform the coefficients through the following expression:
\begin{equation}
a_{i}(t)=b_{i}(t)e^{-i(\varepsilon _{i}+\omega c^{+}c)t}  
\label{coeff}
\end{equation}
and, expanding the amplitudes $b_{i}(t)$ in photon number states of the
cavity, we have
\begin{equation}
\bigskip b_{i}(t)=\sum_{n} b_{i,n}(t)\left| n\right\rangle .
\label{coeff2}
\end{equation}
Thus, in case of exact resonance we have the following equations for the
coefficients of transformation: 
\begin{equation}
i\frac{d}{dt} B_{n}(t) = {\bf W} (t)B_{n}(t),  
\label{transform}
\end{equation}
where $B_{n}(t)$ is a column of coefficients $b_{in}(t)$, and

\begin{eqnarray}
{\bf W}(t) \, {\Huge=} \, \left( 
\begin{array}{cccc}
0&W_{1}(t)&0&0\\
W_{1}(t)&0&W_{2}(t)&\sqrt{n+1}\beta(t)\\ 
0&W_{2}(t)&0&0\\ 
0&\sqrt{n+1}\beta(t)&0&0
\end{array}
\right)  
\label{W}
\end{eqnarray}
Here the matrix elements $W_{1}(t),W_{2}(t),\beta (t)$ are assumed to be
real-valued, since any constant phases of laser fields can be absorbed in
the definition of amplitudes $b_{in}(t)$.

We are interested in solutions of the equation (\ref{transform}), (\ref{W}) in
adiabatic approximation. Two adiabatic energy eigenstates are degenerate,
with null eigenvalue:
\smallskip
\begin{eqnarray}
\Phi_{1}(t)=\left[ 
\begin{array} {c}
\cos \vartheta (t) \\ 
0 \\ 
0 \\
-\sin \vartheta (t) 
\end{array}
\right] ,\qquad \quad \Phi_{2}(t)=\left[ 
\begin{array}{c}
\sin \varphi (t)\sin \vartheta (t) \\ 
0 \\ 
- \cos \varphi (t)\\
\sin \varphi (t)\cos \vartheta (t)
\end{array}
\right] ,\label{eigenvectors1} 
\end{eqnarray}
\begin{equation}
\lambda _{1,2}(t)=0  
\label{eigenvalues1}
\end{equation}
\smallskip
The remaining eigenvectors and eigenvalues are:
\smallskip
\begin{equation}
\Phi_{3}(t)=\frac{1}{\sqrt{2}}\left[ 
\begin{array}{c}
\cos \varphi (t)\sin \vartheta (t) \\ 
1 \\ 
\sin \varphi (t) \\
\cos \varphi (t)\cos \vartheta (t)
\end{array}
\right] ,\qquad  \quad \Phi_{4}(t)=\frac{1}{\sqrt{2}}%
\left[ 
\begin{array}{c}
\cos \varphi (t)\sin \vartheta (t) \\ 
-1 \\ 
\sin \varphi (t) \\
\cos \varphi (t)\cos \vartheta (t)
\end{array}
\right] ,
\label{eigenvectors2}
\end{equation}

\begin{equation}
\lambda _{3}(t)=+\Omega (t),\qquad \lambda _{4}(t)=-\Omega (t)  
\label{eigenvalues2}
\end{equation}
\smallskip
where $\Omega (t)=\sqrt{W_{1}(t)^{2}+W_{2}(t)^{2}+(n+1)\beta
(t)^{2}}$. $\vartheta (t)$ and $\varphi (t)$ are defined by
\smallskip
\begin{equation}
\tan \vartheta (t)=%
%TCIMACRO{\dfrac{W_{1}(t)}{\sqrt{n+1}\beta (t)} }
%BeginExpansion
{\displaystyle {W_{1}(t) \over \sqrt{n+1}\beta (t)}}%
%EndExpansion
,\quad \tan \varphi (t)= 
%TCIMACRO{
%\dfrac{W_{2}(t)}{\sqrt{V_{1}(t)^{2}+V_{2}(t)^{2}+(n+1)\beta (t)^{2}}} }
%BeginExpansion
{\displaystyle {W_{2}(t) \over \sqrt{W_{1}(t)^{2}+(n+1)\beta (t)^{2}}}}%
%EndExpansion
.  \label{angles}
\end{equation}
It is easily seen that the degenerate eigenvectors $\Phi_{1,2}(t)$
(\ref{eigenvectors1}) receive no contribution from the excited atomic state
$\left|2\right\rangle$. These states are therefore termed {\it trapped} or
{\it dark} states. The significant feature of our scheme is that, if the
system remains in some superposition of the two degenerate dark states,
atomic spontaneous emission plays no role in the system dynamics.

\section{GENERATION OF CAVITY SUPERPOSITION STATES}
In order to place diagonally the matrix ${\bf W}(t)$ (\ref{W}) we introduce a
unitary matrix 
\begin{equation}
{\bf U}(t)\,{\huge =}\,\left( 
\begin{array}{cccc}
\cos \vartheta (t) & \sin \vartheta (t)\sin \varphi (t) & \frac{1}{\sqrt{2}}%
\sin \vartheta (t)\cos \varphi (t) & \frac{1}{\sqrt{2}}\sin \vartheta
(t)\cos \varphi (t) \\ 
0 & 0 & \frac{1}{\sqrt{2}} & -\frac{1}{\sqrt{2}} \\ 
0 & -\cos \varphi (t) & \frac{1}{\sqrt{2}}\sin \varphi (t) & \frac{1}{\sqrt{2%
}}\sin \varphi (t) \\ 
-\sin \vartheta (t) & \cos \vartheta (t)\sin \varphi (t) & \frac{1}{\sqrt{2}}%
\cos \vartheta (t)\cos \varphi (t) & \frac{1}{\sqrt{2}}\cos \vartheta
(t)\cos \varphi (t)
\end{array}
\right)  
\label{U}
\end{equation}
Taking into consideration the non-adiabatic interactions between adiabatic
energy eigenstates (\ref{eigenvectors1}), (\ref{eigenvectors2}) we obtain:
\begin{equation}
i\frac{dC_{n}(t)}{dt}=\widetilde{W}_{n}(t)C_{n}(t),  \label{Schrodinger1}
\end{equation}
where
\begin{equation}
\widetilde{W}(t)=U^{-1}(t)W(t)U(t)+i\frac{dU^{-1}(t)}{dt}U(t),  \label{W1}
\end{equation}
\begin{equation}
B_{n}(t)=U(t)C_{n}(t).  \label{transform1}
\end{equation}
Obviously, through the conditions
\begin{eqnarray}
\frac{d\vartheta (t)}{dt}\ll\Omega (t)
\label{condition1} \\
\bigskip
\bigskip
\frac{d\varphi (t)}{dt}\ll\Omega (t),
\label{condition2}
\end{eqnarray}
which imply that the effective pulse area is very large
\begin{equation}
S^{\prime }=\int\limits_{-\infty}^{+\infty} d\tau \Omega(\tau ) \gg 1,
\label{area}
\end{equation}
the coupling of the dark states (\ref{eigenvectors1}) to the bright states
(\ref{eigenvectors2}) can be disregarded. Hence, the matrix
$\widetilde{W}_{n}(t)$ is given by
\begin{equation}
\widetilde{W}_{n}(t)\approx \left( 
\begin{array}{cccc}
0 & -i\stackrel{\cdot }{\vartheta }\sin \varphi (t) & 0 & 0 \\ 
i\stackrel{\cdot }{\vartheta }\sin \varphi (t) & 0 & 0 & 0 \\ 
0 & 0 & \Omega (t) & 0 \\ 
0 & 0 & 0 & -\Omega (t)
\end{array}
\right)  \label{W2}
\end{equation}
Note that because of the degeneracy the coupling between dark states
(\ref{eigenvectors1}) still exists.
The solution of the equation (\ref{Schrodinger1}) with the matrix (\ref{W2})
has the simple form
\begin{equation}
C_{n}(t)=R(t,-\infty )C_{n}(-\infty )  \label{coeff1}
\end{equation}
Here $C_{n}(-\infty )$ is defined by initial conditions and the evolution
operator is given by
\begin{equation}
{\huge R(t,-\infty )=}\left( 
\begin{array}{cccc}
\cos \gamma (t) & -\sin \gamma (t) & 0 & 0 \\ 
\sin \gamma (t) & \cos \gamma (t) & 0 & 0 \\ 
0 & 0 & e^{-iS(t)} & 0 \\ 
0 & 0 & 0 & e^{iS(t)}
\end{array}
\right) , \label{R}
\end{equation}
with
\begin{equation}
\gamma (t)=\int\limits_{-\infty}^{t} \stackrel{\cdot }{\vartheta }(t)\sin \varphi (t)dt
\label{tetta}
\end{equation}
\smallskip
\begin{equation}
%S(t)=\stackrel{t}{\stackunder{-\infty }{\int }\mathrel{\mathop{\int
%}\limits_{-\infty }}}\Omega (t)dt  \label{S}
S(t)=\int\limits_{-\infty}^{t} \Omega (\tau)d\tau  \label{S}
\end{equation}
Proceed from (\ref{transform1}) and (\ref{coeff1}) we obtain the general
solutions for coefficients
\begin{equation}
B_{n}(t)=U(t)R(t,-\infty )U^{-1}(-\infty )B_{n}(-\infty )  
\label{transform2}
\end{equation}
The atom is supposed to be initially in the state $\left|1\right\rangle$
\begin{equation}
B_{n}(-\infty )=\left( 
\begin{array}{c}
1 \\ 
0 \\ 
0 \\ 
0
\end{array}
\right)  
\label{initial1}
\end{equation}
Thus, from (\ref{transform2}) we obtain the solutions for initial condition (\ref{initial1}):
\begin{equation}
{\Huge B}_{n}{\Huge (t)=}\left( 
\begin{array}{c}
\cos \vartheta (t)\cos \gamma (t)+\sin \vartheta (t)\sin \varphi (t) \\ 
0 \\ 
-\cos \varphi (t)\sin \gamma (t) \\ 
-\sin \vartheta (t)\cos \gamma (t)+\cos \vartheta (t)\sin \varphi (t)\sin
\gamma (t)
\end{array}
\right)  
\label{final}
\end{equation}
Our scheme requires the passage of the atom through the cavity field and
laser field $V_{1}(t)$. After that, the atom interacts with the second laser
field $V_{2}(t)$. Then, we consider the case, when laser field $%
V_{2}(t) $ turn off after the cavity quantum field and laser field $V_{1}(t)$
cease. To achieve this pulse sequence we choose
\begin{equation}
\vartheta (+\infty )=0,\qquad \varphi (+\infty )=\frac{\pi }{2}  
\label{angles1}
\end{equation}
The result of this sequence of interactions is the state
\begin{equation}
B_{n}(+\infty )=\left( 
\begin{array}{c}
\cos \gamma \\ 
0 \\ 
0 \\ 
\sin \gamma
\end{array}
\right)  
\label{final1}
\end{equation}
As easily can be seen from the superposition expressions (\ref{basis}), (\ref
{coeff}) (\ref{coeff2}) and (\ref{final1}), in case of formation of dark
states, and small non-adiabatic interaction of degenerate dressed states, the
asymptotic wave function for $t\longrightarrow +\infty $ can be written as
\begin{equation}
\left| \Psi \right\rangle\ =e^{-i(\varepsilon _{1}+n\omega )t} \,\left[\, \cos \gamma \left|
n\right\rangle \left| \psi _{1}\right\rangle \, + \, e^{i\varphi _{1}}
e^{-i\omega_{1}t} \sin \gamma \left| n+1\right\rangle  \left| \psi _{4}\right\rangle \right ]   
\label{final2}
\end{equation}
here $\varphi _{1}$ is the constant phase factor of the laser field $V_{1}(t)$, and
\begin{equation}
\gamma =\int\limits_{-\infty}^{+\infty} \stackrel{\cdot }{\vartheta }(t)\sin \varphi (t)dt
\label{tetta2}
\end{equation}
Especially, if the cavity-mode is initially in the vacuum state,
i.e. 0-photon Fock state, the resulting wave function is given by a
superposition of vacuum and 1-photon states:
\begin{equation}
\Phi (t)=e^{-i\varepsilon _{1}t} \left [ \cos \gamma \left| 0\right\rangle \psi
_{1}+e^{i\varphi _{1}}\sin \gamma e^{-i\omega _{1}t}\left| 1\right\rangle
\psi _{4} \right ],  
\label{final3}
\end{equation}
One can achieve the desired superposition just by changing the relative delay
between pulses.

Fig.~\ref{solut} shows an example of population histories for this sequence
of interaction of three Gaussian pulses. Fig.~\ref{gamma} demonstrates the
dependence of mixing angle $\gamma$ on the delay between pulses calculated numerically.

We note that the only conditions for successful operation (\ref{condition1}),
(\ref{condition2}) can be easily fulfilled. The scheme is robust against
small variations of parameters.

\section{ENTANGLEMENT OF TWO FOUR-LEVEL ATOMS}
In the following we modify the scheme presented above to prepare an
entanglement of atomic coherent states. Let us consider now two four-level
atoms and a configuration of interaction fields when the laser pulses $V_1
(t)$ and $V_2 (t)$ equal each other (Fig.~\ref{entang-sch}). We assume, that initially ($t \to
- \infty$) the atom named as first is in $\left| \psi_1
\right\rangle $ state, the second one is in $\left| \psi_4 \right\rangle
$ state, and the cavity field is in the vacuum state $\left| n=0\right\rangle $:

\begin{equation}
\left| \psi_1 \right\rangle_1 \otimes \left| \psi_4 \right\rangle_2 \otimes
\left| n=0 \right\rangle
\label {entin}
\end{equation}

At the first step the first atom interacts with quantum field and then with
the laser field (see Fig.~\ref{entang}). The initial state asymptotically
coincides with the dressed state $\Phi _1 (t)$ (\ref{eigenvectors1}) at very
early times. At the end of the interaction, as can be easily seen from
(\ref{eigenvectors1}), (\ref{eigenvectors2}) the following connections exist
between adiabatic and bare states.

\begin{equation}
\Phi _1 \to -\left| \psi_4 \right\rangle _1 \otimes \left|n=1\right\rangle 
\label {ent2}
\end{equation}
and
\begin{equation}
\Phi _2 \to \frac{1}{\sqrt{2}}\left( \left|\psi_1 \right\rangle _1 -
  \left|\psi_3 \right\rangle _1 \right) \otimes \left|n=0\right\rangle
\label {ent3}
\end{equation}
 
Again, under the conditions (\ref{condition1}), (\ref{condition2}) in the
adiabatic limit, we must take into account only transitions between the
degenerate dressed states $\Phi _1 (t)$, $\Phi _2 (t)$. 

At the next step, the reverse sequence of pulses (i.e. the equal laser fields
then the cavity field) acts on the second atom, which is still in
$\left| \psi_4 \right\rangle _2$ state (see Fig.~\ref{entang2}). It is clear, that the result of this
interaction depends on the number of photons in the cavity. If the cavity is
in the zero-photon state $\left|n=0\right\rangle$ as (\ref{ent3}), the second
atom remains in the $\left| \psi_4 \right\rangle _2$ state. It is straightforward to verify that for one-photon cavity state
$\left|n=1\right\rangle$ (as in (\ref{ent2})) the result of this sequence of
interactions is:

\begin{equation}
\frac{1}{\sqrt{2}} \left( \left| \psi_1 \right\rangle _2 - \left| \psi_3
  \right\rangle _2 \right) \otimes \left|n=0\right\rangle 
\label {ent4}
\end{equation}

So, the initial state (\ref{entin}) turns into the entanglement state

\begin{equation}
\left| \psi_1 \right\rangle _1 \otimes \left| \psi_4 \right\rangle _2 \otimes
\left| n=0\right\rangle \to \Big( \frac{1}{2} \left( \left| \psi_1
  \right\rangle _1 - \left| \psi_3 \right\rangle _1 \right) \otimes \left|
  \psi_4 \right\rangle _2 - \frac{1}{2}  \left| \psi_4 \right\rangle _1
\otimes  \left( \left| \psi_1 \right\rangle _2 - \left| \psi_3 \right\rangle
  _2 \right) \Big) \otimes \left|n=0\right\rangle
\label {entanglement}
\end{equation}

Thus, we obtain the entanglement state of two spatially separated atoms.

\section{SUMMARY AND CONCLUSIONS}

In the present paper we have suggested a method to create a quantum
superposition of Fock states in cavity. We obtained the wave function
(\ref{final2}), (\ref{final3}) in case of formation of dark states, and non-adiabatic
interaction of two degenerate ''dressed'' states. As can be seen, the final
result does not depend on the effective area of pulses (\ref{area}), and
depends only on the angle $\gamma $ or on the ratio of delay time and
durations of two laser fields. That makes this method of creation of
coherent superposition effective and robust against small variations of
parameters. Arbitrary coherent superposition of Fock states in the cavity can
be obtained by changing the value of the angle $\gamma $, i.e. the delay
between laser pulses. Especially, in the case of $\theta =\frac{\pi }{2}$,
we obtain Fock states of the cavity field. 

A variation of these basic ideas for interaction of cavity and laser fields
with atomic systems is to employ the particular properties of the cavity
field for the generation of entanglement of atoms. In this arrangement two
atoms initially prepared in $\left| \psi_1 \right\rangle _1$ and  $\left|
  \psi_4 \right\rangle _2$ states, adiabatically interact with direct and
reverse sequences of the cavity field and two equal laser fields. After the
interaction multiparticle entanglement of spatially separated neutral atoms
is formed.

\acknowledgments The author is grateful to Prof. R.G.Unanyan and
Prof. A.D.Gazazyan for helpful discussions and continued support.

\begin{figure}[tbp]
\caption{The scheme of the four-level atomic system.}
\label{fourl}
\end{figure}

\begin{figure}[tbp]
\caption{Sequence of interaction of pulses.}
\label{pulses}
\end{figure}

\begin{figure}[tbp]
\caption{Proposed configuration for the generation of superposition of Fock
  states using four-level atoms. The propagation directions of laser pulses are perpendicular to the page.}
\label{config}
\end{figure}

\begin{figure}[tbp]
\caption{Time evolution of populations obtained from numerical solutions of
  the Schr\"{o}dinger equation. The amplitudes of the pulses are:
  $A_1=A_2=20$,$ A_{\beta}=4 $; pulse lengths are: $T_1=10$, $T_2=20$, $T_{\beta}= 20$;
 the delay between the laser pulses $\tau = 20 $. }
\label{solut}
\end{figure}

\begin{figure}[tbp]
\caption{Mixing angle $\gamma$ as a function of delay between pulses.}
\label{gamma}
\end{figure}

\begin{figure}[tbp]
\caption{Schematics of the preparation of entanglement of two four-level atoms.}
\label{entang-sch}
\end{figure}

\begin{figure}[tbp]
\caption{Generation of entanglement of atomic coherent states. Sequence of interaction 
of pulses for the first atom.}
\label{entang}
\end{figure}

\begin{figure}[tbp]
\caption{Generation of entanglement of atomic coherent states. Sequence of interaction 
of pulses for the second atom.}
\label{entang2}
\end{figure}

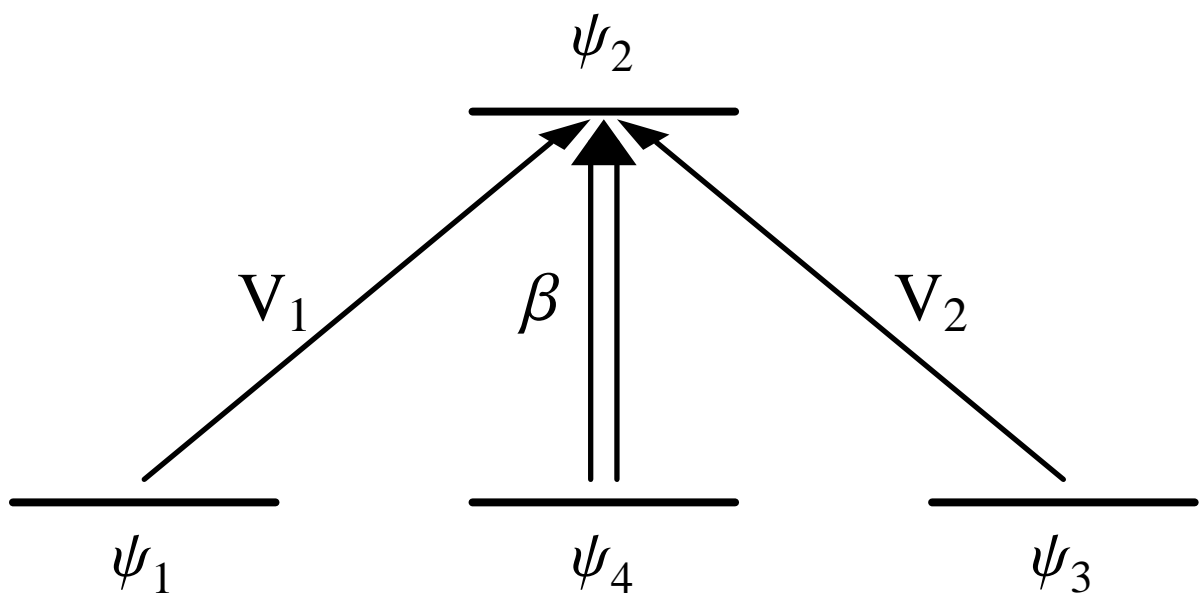
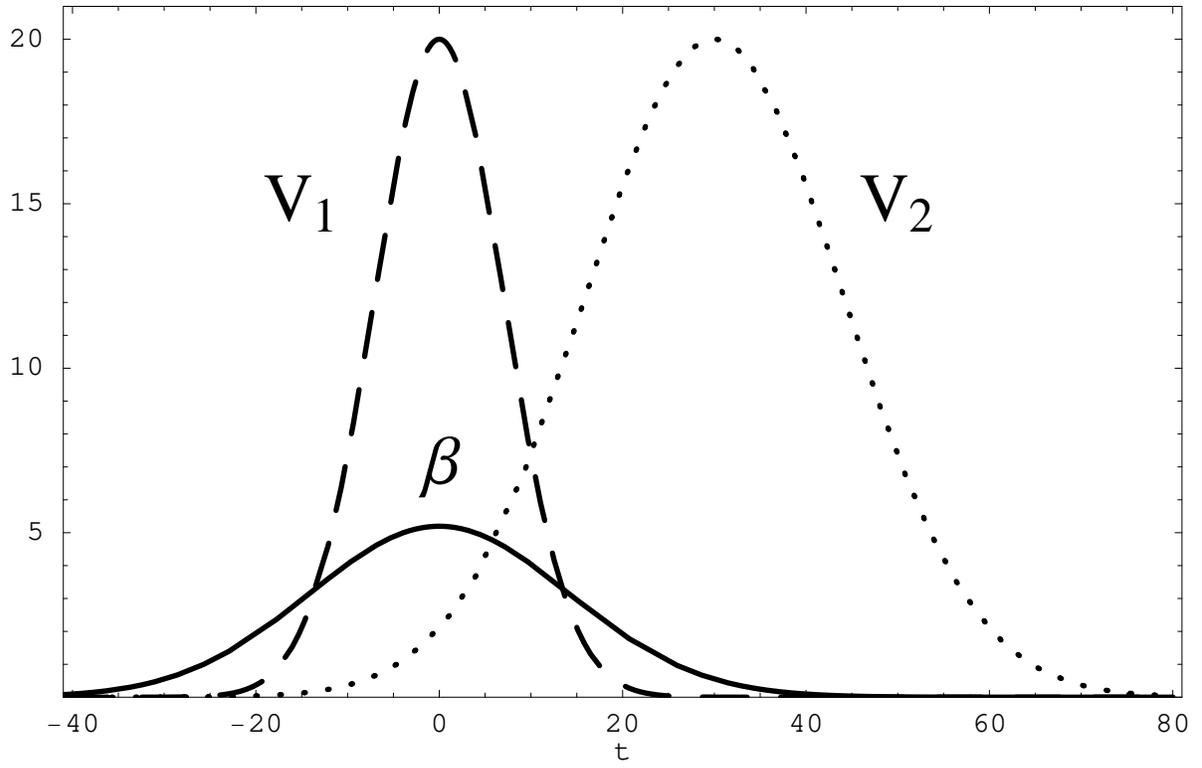
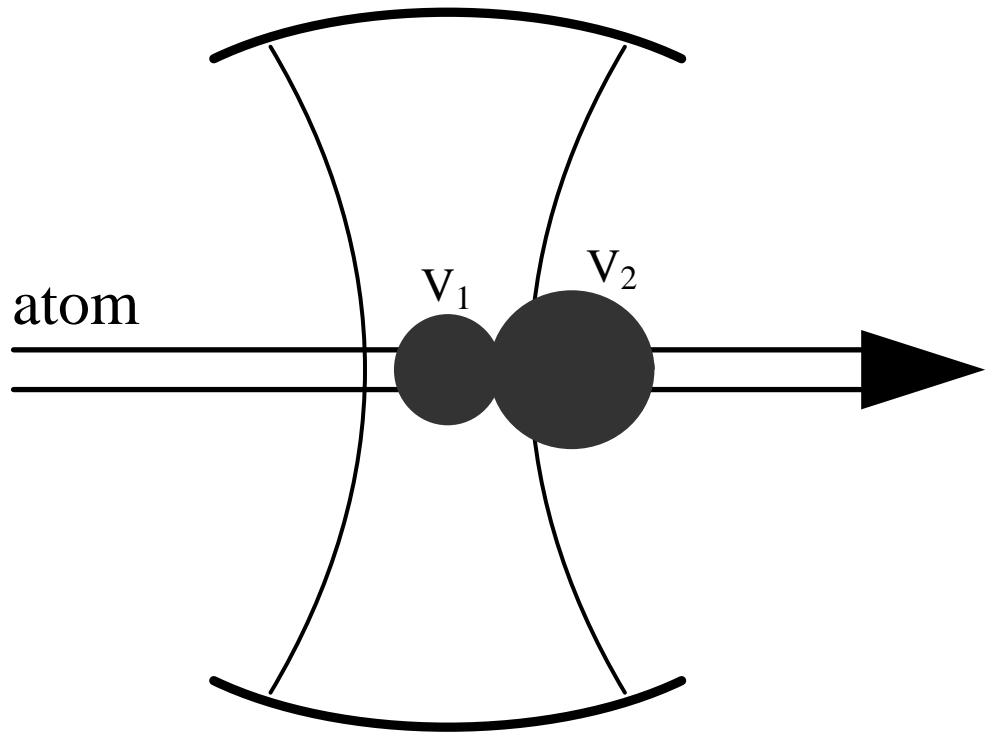
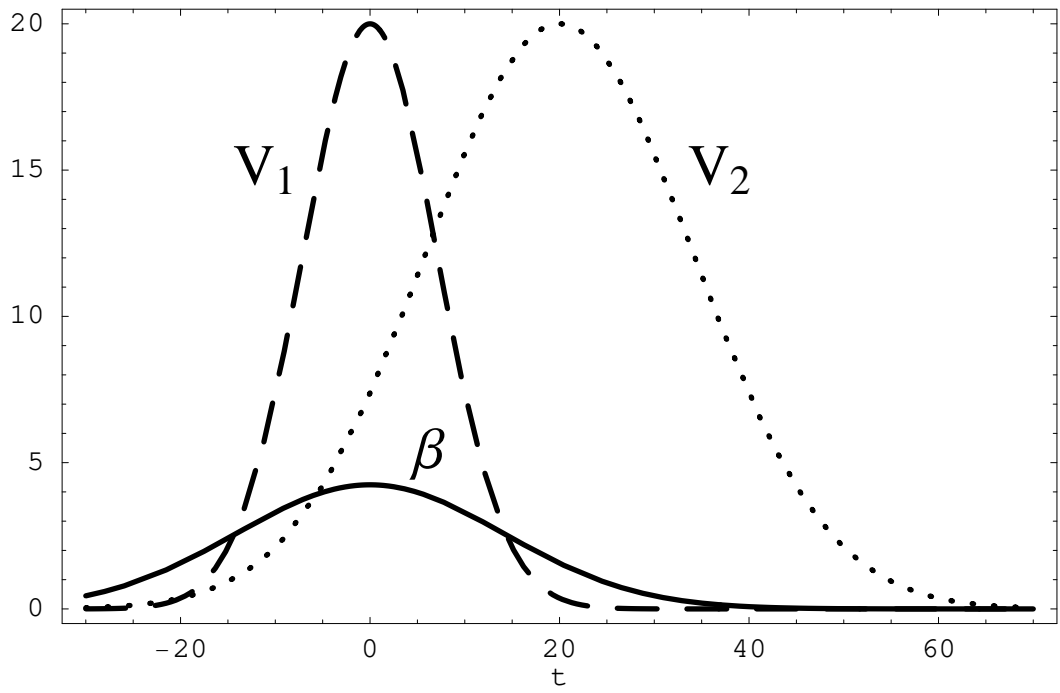
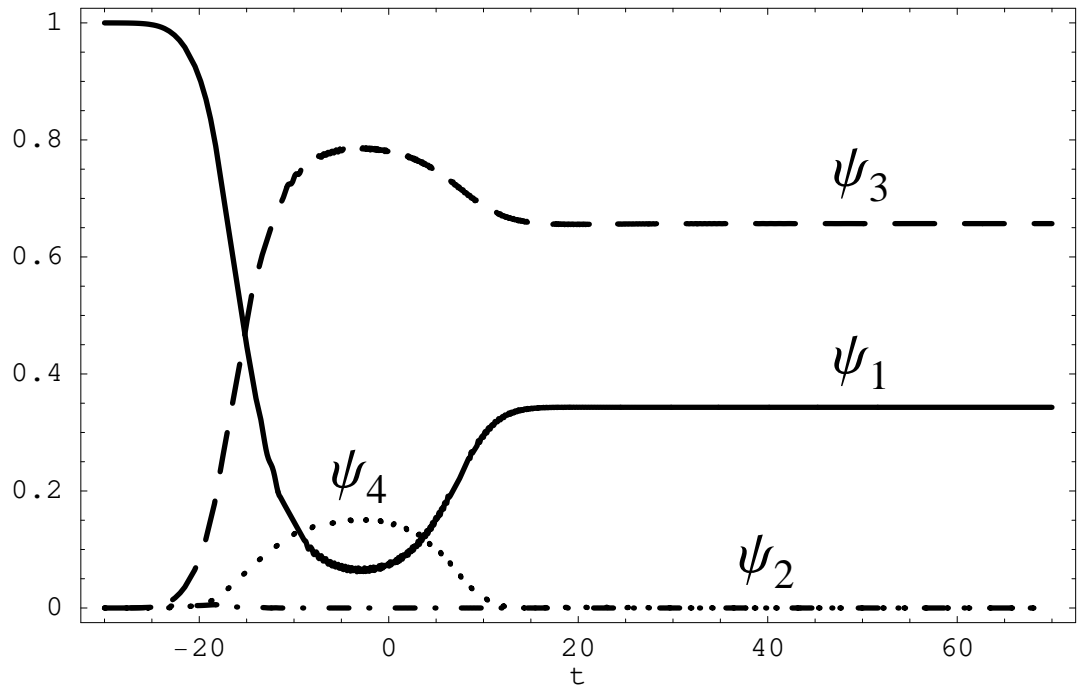
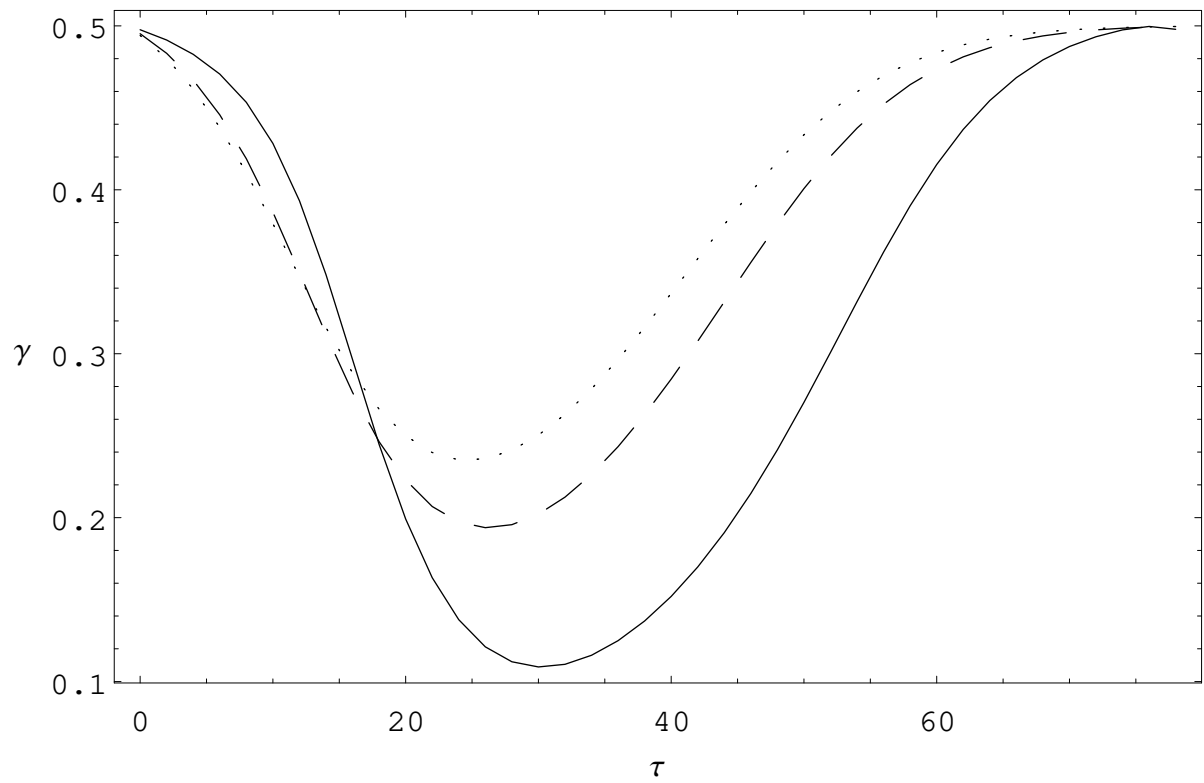
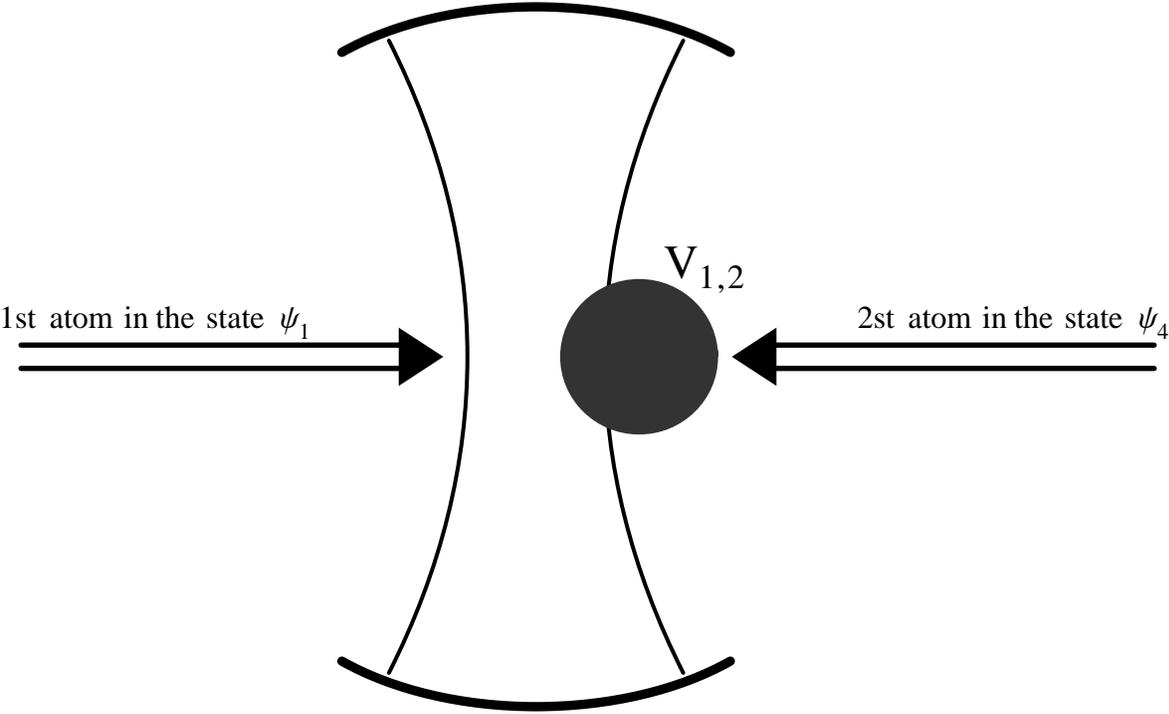
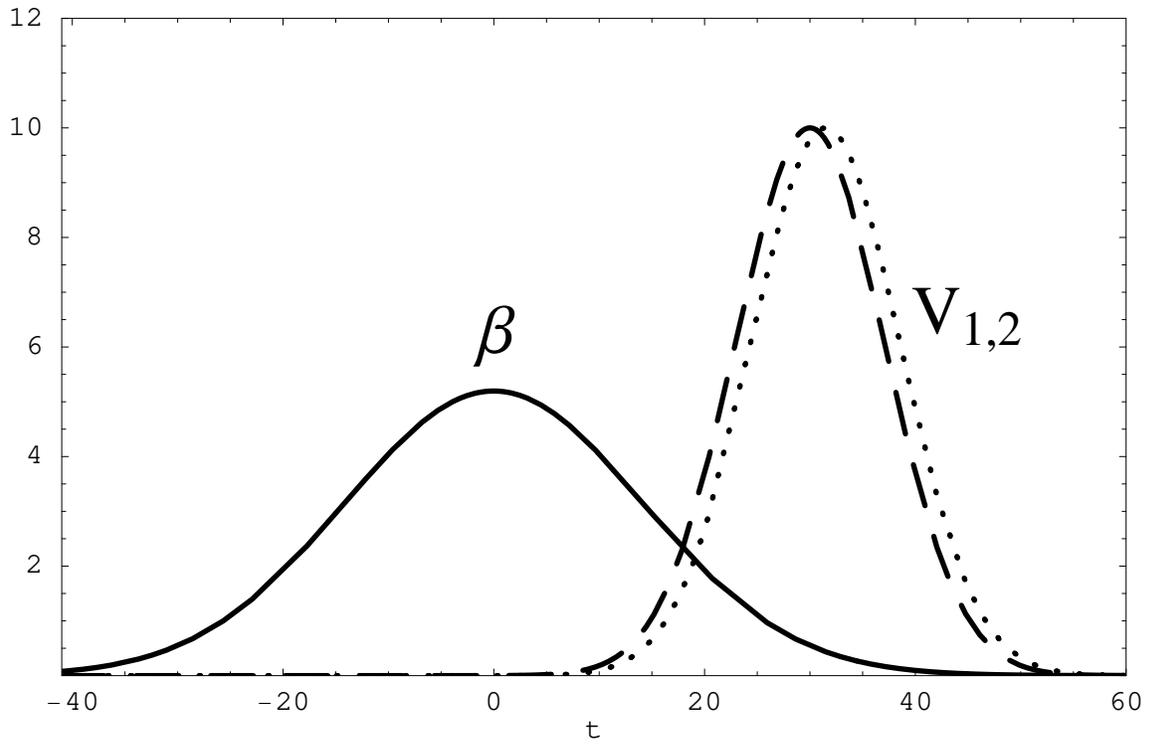
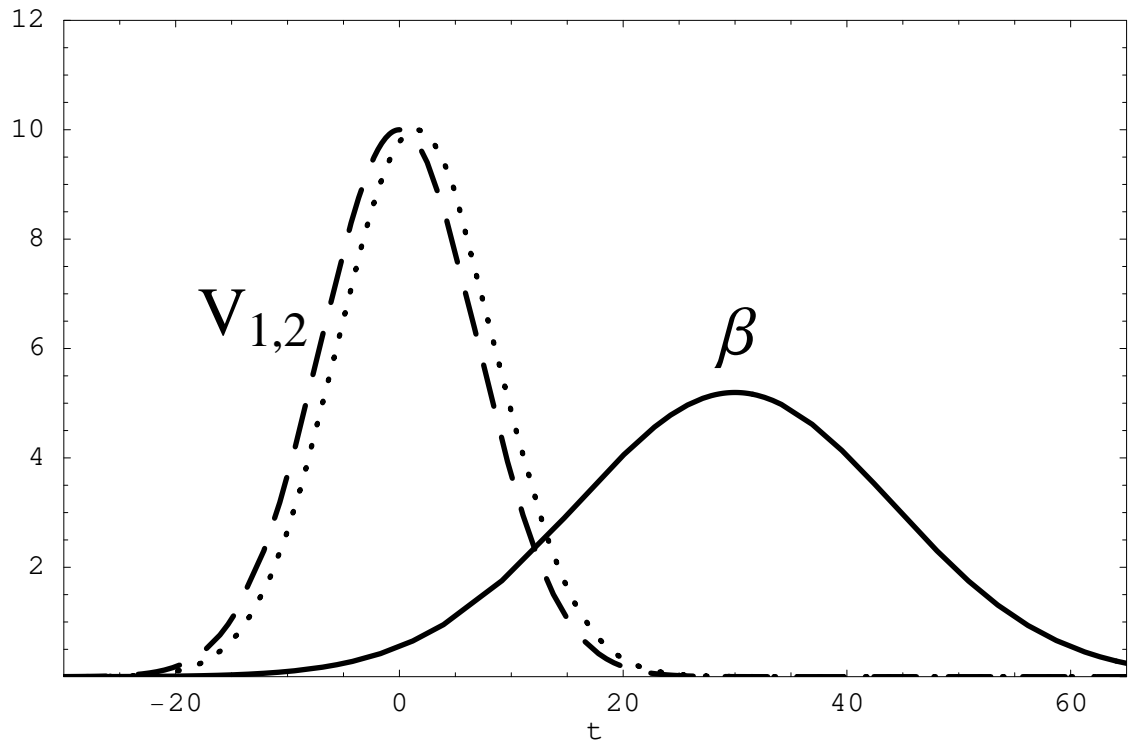
\end{document}